\newcommand\tit[1]{``#1,''}
\newcommand{\C}{{\Bbb C}}
\newcommand{\Hil}{{\cal H}}
\newcommand{\R}{{\Bbb R}}
\newcommand{\Cl}{{\frak Cl}}
\newcommand{\Id}{\openone}
\newcommand{\G}{{\mit\Gamma}}
\newcommand{\un}{{\frak u}}
\newcommand{\gl}{{\frak gl}}
\newcommand{\T}{{\Bbb T}}
\newcommand{\Tnl}{\T^{2n}_{1/l}}
\newcommand{\eq}[1]{Eq.~(\ref{#1})}
\newcommand{\eqs}[2]{Eqs.~(\ref{#1}, \ref{#2})}
\newcommand{\Mat}[1]{\C(#1{\times}#1)}
\newcommand{\mod}[1]{({\rm mod}~#1)}
\newcommand{\ad}{{\rm ad}}
\newcommand{\zrel}{\to_\zeta}
\newcommand{\zdow}{\downarrow_\zeta}
\newcommand{\ie}{{\it i.e.}, }

\def\iso{\cong}
\def\hc{\dag}

\documentstyle[aps,prb,eqsecnum,amssymb,amsbsy]{revtex}
\begin{document}
\draft
\makeatletter
\def\section{\@mainheadtrue
\@startsection {section}{1}{\z@}{0.8cm plus1ex minus
 .2ex}{0.5cm plus1ex minus.2ex}{\reset@font\small\bf\sffamily\flushleft}}
\makeatother
\pagestyle{myheadings}
\markright{{\hfil\sf A. Yu. Vlasov: Noncommutative tori and quantum gates}}
\addtolength{\topmargin}{1cm}
\title{\sf Noncommutative tori and universal sets of non-binary quantum gates}
\author{Alexander Yu. Vlasov\thanks{
 E-mail: {\tt qubeat@mail.ru} or {\tt alex@protection.spb.su}}}
\address{Federal Radiological Center (IRH),
197101, Mira Street 8, St.-Petersburg, Russia}
\date{v1: 1 Dec 2000 $\approx$ v2: 18 Jun 2002}
\maketitle
\begin{abstract}
A problem of universality in simulation of evolution of quantum system and
in theory of quantum computations is related with the possibility of expression
or approximation of arbitrary unitary transformation by composition
of specific unitary transformations (quantum gates) from given set.
In an earlier paper (Ref.\onlinecite{VlaUnCl}) application of Clifford algebras
to constructions of universal sets of binary quantum gates $U_k \in U(2^n)$
was shown. For application
of a similar approach to non-binary quantum gates $U_k \in U(l^n)$
in present work is used rational noncommutative torus ${\Bbb T}^{2n}_{1/l}$.
A set of universal non-binary two-gates is presented here as one example.
\end{abstract}

\section{Introduction}

Let $\Hil_l$ be a Hilbert space of quantum system with $l$ states
and $\Hil_l^n \equiv \Hil_l^{\otimes n}$ is an $l^n$-dimensional Hilbert
space of $n$ systems expressed as $n$-th tensor power. For $l=2$
element of $\Hil_2$ ($\Hil_2^n$) is usually called a qubit(s).
An algebra $\Mat{l^n}$ of all complex $l^n{\times}l^n$ matrices corresponds
to general linear transformations of $\Hil_l^n$ and a group of
unitary matrices $U(l^n)$ corresponds to physically possible
evolution. Because of the natural structure of tensor power it is possible
to consider groups of transformations of subsystems
$U(l^k) \iso U(l^n) \cap \bigl(\Mat{l^k} \otimes \Id_l^{\otimes n-k}\bigr)$.
Such transformations correspond to {\em quantum gates}. For $l=2$ they are
usually called $k$-qubits gates.

The problem of universality in quantum simulation and computation is related
to approximation with necessary precision (in some appropriate norm) of
arbitrary unitary transformation $U \in U(l^n)$ of $\Hil_l^n$ as a product
of matrices $U_k$ from some fixed set called here {\em the universal set of
quantum gates}. One origin of the task was the idea of generalization
of the Church--Turing principle from computer science to physical systems in
works by David Deutsch, \cite{DeuTur,DeuGate} where it was suggested there be
some universal set of matrices for ``{\em binary}'' quantum gates with $l=2$.

 It was found also that it is possible to express necessary conditions
of universality by using elegant framework with Lie algebra $\un(2^n)$ of
Lie group $\mathrm U(2^n)$. \cite{DPDV2,DeuUn,UnSim} In the approach it is
necessary to find a set of elements $A_k \in \un(2^n)$, $A_k^\hc = -A_k$,
which generate full algebra $\un(2^n)$. It is possible to use
$U^{\tau}_k = \exp(\tau A_k)$ with infinitesimal
parameter $\tau \in \R$ as a universal set of quantum gates. In a more
physical picture, $A_k \corresponds i H_k$, where $H_k$ are Hamiltonians
and Lie brackets also contain multipliers with imaginary unit.

 Previous work \cite{VlaUnCl} suggested construction of the universal
set by inclusion $\un(2^n)$ in Clifford algebra $\Cl(2n,\C)\iso\Mat{2^n}$
with Lie algebra structure due to bracket operation
$[a,b] \equiv ab-ba$ ({\em cf.}\ Ref.\onlinecite{ClDir}).
Because commutation laws for basis elements of Clifford algebra are simple
enough due to canonical relations between generators,
\begin{equation}
\G_j\G_k+\G_k\G_j = 2\delta_{kj}\Id,
\label{ClifDef}
\end{equation}
it was possible to represent useful constructions of universal sets with
$2n+1$ elements (see Ref.~\onlinecite{VlaUnCl}).

 For generalization of the method for ``non-binary'' quantum gates
$U(l^n)$, $l > 2$, it is useful to find some algebra with simple commutation
rules, like \eq{ClifDef}, and use it to express elements of $\un(l^n)$.
In this paper as such generalization of $\Cl(2n,\C) \iso \Mat{2^n}$
is used {\em noncommutative torus} $\Tnl \iso \Mat{l^n}$
with $2n$ generators:
\begin{mathletters}
\label{NCTorDef}
\begin{eqnarray}
& (T_k)^l = \Id, \\
& T_j T_k = \zeta T_k T_j ~~ (j < k), \\
&\zeta = \exp(2\pi i/l).
\end{eqnarray}
\end{mathletters}
For $l=2$ \eq{NCTorDef} coincides with \eq{ClifDef}. For $l > 2$, elements
$T_k \notin \un(l^n)$ for most or all $k$, but it is possible to use
representation $\un(N) \subset \Re\bigl(\gl(N,\C)\bigr)$. More
concretely, it is enough to find a set of elements $M_k \in \gl(N,\C)$,
which generate full algebra $\gl(N,\C)$. It is possible to use
$G^{\tau}_k = \exp\bigl(i\tau(M_k + M_k^\hc)\bigr)$,
$F^{\tau}_k = \exp\bigl(\tau(M_k - M_k^\hc)\bigr)$ with infinitesimal
parameter $\tau \in \R$ as a universal set of quantum gates.

\section{Noncommutative torus $\pmb{\T^2_\theta}$ and quantum one-gates}

Let us consider one-particle transformations. For two-dimensional
case, $\Hil_2$,
any two Pauli matrices, for example $\sigma_x$ and $\sigma_z$, generate full
four-dimensional basis of $\Mat 2$, \ie $\{\sigma_x^2=\sigma_y^2 = \Id_2,
\sigma_x, \sigma_z, \sigma_y = i\sigma_x\sigma_z\}$.

Analogously, two generators $U,V$ of noncommutative torus $\T^2_\theta$
defined as\cite{ConnesNG}
\begin{equation}
 U V = \exp(2 \pi i \theta) V U,\quad V V^\hc = U U^\hc = \Id,
\label{T2Def}
\end{equation}
produce for rational $\theta=1/l$ an algebra isomorphic to $\Mat l$. The
basis of the algebra are $l^2$ elements $U^m V^n$, $m,n = 0,\ldots,l-1$.

Let us use the Weyl representation of $U$ and $V$ as the right cyclic shift
operator and its Fourier transform:
\begin{equation}
 U_{kj} = \delta_{k+1 \mod l,j}, \quad
 V_{kj} = \exp(2 \pi i k/l) \delta_{kj}.
\label{defUV}
\end{equation}
The representation and $U^m V^n$ basis are well known in quantum information
science after application to the theory of quantum error correction.\cite{QEC}

To find transformation between basis $U^m V^n$ and canonical basis
$E^{ab}$ of $\Mat l$, there $(E^{ab})_{jk} = \delta_{aj}\delta_{bk}$,
it is enough to use $E^{00}=1/l\sum_{k=0}^{l-1}V^k$ together with
$E^{ab}=U^{l-a}E^{00}U^b$.

Let us show that any $U^m V^n$ (except $\Id_l$ for $m=n=0$) can be generated
from $U$ and $V$ using only commutators
$[A,B] \equiv (\ad~A)\, B \equiv A B - B A$.
For $U$ and $V$ commutator is simply $[U,V] = (1-\zeta)\, UV \propto UV$,
where $\zeta = \exp(2 \pi i / l)$.
It is convenient to use ``$\ad$'' for consecutive commutators, for example
$(\ad~A)^2 B \equiv \bigl[A,[A,B]\bigr]$, and symbol ``proportional,''
$A \propto B \Rightarrow A = \alpha B$, to avoid unessential nonzero
complex multipliers $\alpha$.

 Direct expressions for $l^2-1$ elements $U^m V^n$ are
\begin{equation}
U^m V^n \propto (\ad\,U)^{m-1 \mod l} \bigl((\ad\,V)^{n-1 \mod l} [U,V]\bigr)
\label{UmVnUV}
\end{equation}
where $0 \le m, n < l$ are any pair of numbers except \mbox{$m=n=0$},
$(\ad~U)^0$ or $(\ad~V)^0$ corresponds to
absence of the term, and $-1 \mod l = l-1$.

Of course it is possible to suggest simpler expression for
particular values of $m$ and $n$, but \eq{UmVnUV} shows also
application of a third element $W \propto UV$:
\begin{equation}
 U W = \zeta W U,\quad W V = \zeta V W.
\label{propW}
\end{equation}
It is convenient to define
\begin{equation}
W = \zeta^{(l-1)/2} UV, \quad W^l=\Id_l.
\label{defW}
\end{equation}

Similar with case $l=2$ with $(\sigma_x, \sigma_y, \sigma_z)$, any pair
between $(U, V, W)$ may be used for generation of a full algebra due to
\eq{UmVnUV} together with possibility to express initial pair $(U, V)$
from $(U, W)$ or $(V, W)$:
\begin{equation}
  V \propto (\ad\,U)^{l-1} W, \quad U \propto (\ad\,V)^{l-1} W,
\end{equation}
but for $l \ge 3$ there is a special property
that should be taken into account. Let us use
notation $A{\zrel}B$ for $A B = \zeta B A$.
The definition of relation ``$\zrel$'' is asymmetric for $l \ge 3$.

{\samepage
It is clear from the diagram:
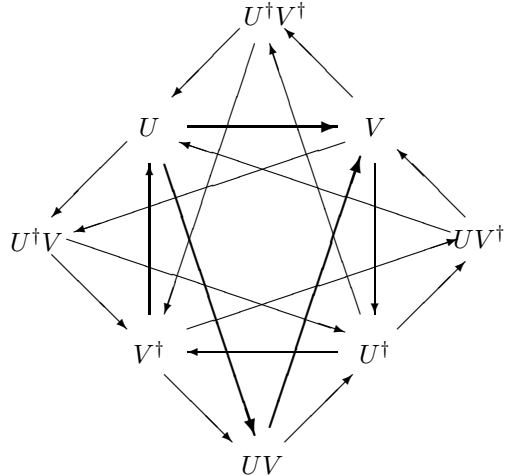
\begin{center}
\unitlength=1mm
\begin{picture}(65.00,65.00)
\thicklines
\put(25.00,50.00){\vector(1,0){20.00}}
\put(22.00,45.00){\vector(1,-3){12.00}}
\put(36.00,10.00){\vector(1,3){12.00}}
\thinlines
\put(5.00,35.00){\makebox(0,0)[cc]{$U^\hc V$}}
\put(35.00,5.00){\makebox(0,0)[cc]{$UV$}}
\put(65.00,35.00){\makebox(0,0)[cc]{$UV^\hc $ ,}}
\put(35.00,65.00){\makebox(0,0)[cc]{$~~~U^\hc V^\hc$}}
\put(20.00,50.00){\makebox(0,0)[cc]{$U$}}
\put(50.00,50.00){\makebox(0,0)[cc]{$V$}}
\put(20.00,20.00){\makebox(0,0)[cc]{$V^\hc$}}
\put(50.00,20.00){\makebox(0,0)[cc]{$U^\hc$}}
\put(45.00,20.00){\vector(-1,0){20.00}}
\put(20.00,25.00){\vector(0,1){20.00}}
\put(50.00,45.00){\vector(0,-1){20.00}}
\put(32.00,63.00){\vector(-1,-1){9.00}}
\put(47.00,54.00){\vector(-1,1){9.00}}
\put(22.00,17.00){\vector(1,-1){9.00}}
\put(38.00,8.00){\vector(1,1){9.00}}
\put(34.00,61.00){\vector(-1,-3){12.00}}
\put(48.00,25.00){\vector(-1,3){12.00}}
\put(17.00,48.00){\vector(-1,-1){10.00}}
\put(7.00,33.00){\vector(1,-1){10.00}}
\put(62.00,38.00){\vector(-1,1){9.00}}
\put(53.00,23.00){\vector(1,1){9.00}}
\put(9.00,35.00){\vector(3,-1){37.00}}
\put(46.00,48.00){\vector(-3,-1){36.00}}
\put(60.00,36.00){\vector(-3,1){36.00}}
\put(25.00,23.00){\vector(3,1){36.00}}
\end{picture}
\end{center}}
\noindent that for different sets with three operators the relation may be
transitive or not.

For example, $U \zrel V$ from \eq{T2Def}, $U \zrel W$ and $W \zrel V$
from \eq{propW} and so we have transitive relation $U \zrel W \zrel V$,
\ie {\em ordering}. Let us call it {\em $\zeta$-order} for certainty.
On the other hand, it is simply to check $W^\hc \zrel U$ and $V \zrel W^\hc$
and here is some cyclic graph. The cyclic case is more symmetric, because
all pairs are equivalent.

For the ordered case it is not so, because $\zeta$-order produces a canonical
map to a subset of natural numbers, \ie indexes, and it is convenient for
construction of noncommutative torus $\Tnl$, $\zeta$-ordered by
definition: $\T_k \zrel \T_j$ for $k < j$  [see \eq{NCTorDef}].

Because of the principle here is used the following definition for generators
of $\T^2_{1/l}$:
\begin{equation}
 T_0 \equiv U, \quad T_1 \equiv W.
\label{GenT2}
\end{equation}
where $U$ is defined in \eq{defUV} and $W$ in \eq{defW}.

\section{Representations of noncommutative tori $\pmb{\Tnl}$}

Let us use notation $T_x \equiv U$, $T_y \equiv W$, $T_z = V$,
where $U$, $V$, $W$ are defined in \eqs{defUV}{defW}.
There is $\zeta$-order $T_x \zrel T_y \zrel T_z$, \ie
\begin{equation}
 T_x T_y = \zeta T_y T_x,\quad
 T_y T_z = \zeta T_z T_y,\quad
 T_x T_z = \zeta T_z T_x.
\end{equation}
It is possible
to introduce $2n$ generators of $\T^{2n}_{1/l}$ as
\begin{mathletters}
\label{genT2n}
\begin{eqnarray}
 T_{2k} & = &
 {\underbrace{\Id_l\otimes\cdots\otimes\Id_l}_{n-k-1}}\otimes%
 T_x\otimes\underbrace{T_z\otimes\cdots\otimes T_z}_k \, , \\
 T_{2k+1} & = &
 {\underbrace{\Id_l\otimes\cdots\otimes\Id_l}_{n-k-1}}\otimes%
 T_y\otimes\underbrace{T_z\otimes\cdots\otimes T_z}_k \, ,
\end{eqnarray}
\end{mathletters}
in direct analogy with construction of Clifford algebras.\cite{VlaUnCl,ClDir}

It is clear that different products of $T_k$ generate full
matrix algebra $\Mat{l^n}$, because $T_x$ and $T_y$ generate $\Mat{l}$.
Let us prove that generators \eq{genT2n} satisfy definition
\eq{NCTorDef} of noncommutative torus $\Tnl$.

First, $T_k^l = \Id$ because $T_x^l = T_y^l= T_y^l = \Id_l$.

To prove that $T_k \zrel T_j$ for any $k < j$, it is enough to consider a few
cases (here ``$T$'' means ``any element'' and ``$\zdow$'' marks $\zeta$-order
of only a pair of noncommutative terms in the tensor products):

{\bf Case 1:} $T_{2k} \zrel T_{2k+1}$, $k \ge 0$
{\samepage
\begin{eqnarray*}
T_{2k} =
\smash{\underbrace{\Id_l\otimes\cdots\otimes\Id_l}_{n-k-1}}\otimes %
 &T_x&{}\otimes\smash{\underbrace{T_z\otimes\cdots\otimes T_z}_k} \\
 &\smash{\zdow}&\qquad \\
T_{2k+1} =
 {\Id_l\otimes\cdots\otimes\Id_l}\otimes %
 &T_y&{}\otimes{T_z\otimes\cdots\otimes T_z} \\
\end{eqnarray*}}

{\bf Case 2:} $T_{2k} \zrel T_{2k+j+1}$, $k \ge 0$, $j > 0$
{\samepage
\begin{eqnarray*}
T_{2k} =
\smash{\underbrace{\Id_l\otimes\cdots\otimes\Id_l}_{n-k-1}}\otimes%
 &T_x&{}\otimes\smash{\underbrace{T_z\otimes\cdots\otimes T_z}_k} \\
 &\smash{\zdow}& \\
T_{2k+1+j} =
 {\Id_l\otimes\cdots\otimes T}\otimes%
 &T_z&{}\otimes{T_z\otimes\cdots\otimes T_z} \\
\end{eqnarray*}}

{\bf Case 3:} $T_{2k+1} \zrel T_{2k+1+j}$, $k \ge 0$, $j > 0$
{\samepage
\begin{eqnarray*}
T_{2k+1} =
\smash{\underbrace{\Id_l\otimes\cdots\otimes\Id_l}_{n-k-1}}\otimes%
 &T_y&{}\otimes\smash{\underbrace{T_z\otimes\cdots\otimes T_z}_k} \\
 &\smash{\zdow}& \\
T_{2k+1+j} =
 {\Id_l\otimes\cdots\otimes T}\otimes%
 &T_z&{}\otimes{T_z\otimes\cdots\otimes T_z} \\
\end{eqnarray*}}

\section{Generation of $\pmb{\Tnl}$ by commutators}

Let us prove that for $l>2$ it is possible to generate $\Tnl$
using only commutators of $2n$ elements $T_k$. The case with $l=2$,
$\Cl(2n,\C) \iso \T^{2n}_{1/2}$ was considered in earlier
work,\cite{VlaUnCl} and it was shown that $2n$ generators are not
enough and it is necessary to add any element of third or fourth order.

Here is presented a proof that for $l>2$, $2n$ generators are enough.
Let us instead of $T_i^{n_i}T_j^{n_j}\cdots T_k^{n_k}$ write simply
$T(n_i,n_j,\ldots,n_k)$ if it is possible without lost of clarity.
Sequences of indexes are always chosen ordered $0 \le i<j< \cdots < k <2n$.
Let us use $\#$ for the number of different indexes in product
$\#(n_0,...,n_{k-1}) \equiv k$ and $\Sigma$ for total number
of terms $\Sigma(n_0,...,n_{k-1}) \equiv \sum_{j=0}^{k-1}n_j$.

It is possible to prove proposition using recursion.
The case with $\#=2$ may be expressed by generalization of \eq{UmVnUV}:
\begin{equation}
T(n_i,n_j) \propto
 (\ad\,T_i)^{n_i-1 \mod l} \bigl((\ad\,T_j)^{n_j-1 \mod l} [T_i,T_j]\bigr).
\label{GenTwo}
\end{equation}

Let all cases with $T(n_{i_1},\ldots,n_{i_k})$, $2 \le k < 2n$,
$i_1 < i_2 < \cdots < i_k$, be proved and it is
necessary to generate all $T(n_{i_1},\ldots,n_{i_k},n_j)$
with $i_k < j \le 2n$.

There are a few different cases:
\begin{itemize}
 \item[]{\bf Case 1:} $\Sigma(n_{i_1},\ldots,n_{i_k}) {\rm~mod~} l \ne 0$:
 \begin{equation}
  T(n_{i_1},\ldots,n_{i_k},n_j)
  \propto (\ad\,T_j)^{n_j} T(n_{i_1},\ldots,n_{i_k}).
 \label{cs1}
 \end{equation}
 \item[]{\bf Case 2:} $\Sigma(n_{i_1},\ldots,n_{i_k}) {\rm~mod~} l = 0$
  and \eq{cs1} vanishes.
 \begin{itemize}
  \item[]{\bf Case 2.1:}
   $\exists n_i \in (n_{i_1},\ldots,n_i,\ldots,n_{i_k}), n_i \ne n_j$:
  \begin{equation}
   T(n_{i_1},\ldots,n_{i_k},n_j)
   \propto [T_i,(\ad\,T_j)^{n_j} T(n_{i_1},\ldots,n_i-1,\ldots,n_{i_k})].
  \end{equation}
  \item[]{\bf Case 2.2:}
   $\nexists n_i \in (n_{i_1},\ldots,n_i,\ldots,n_{i_k}), n_i \ne n_j$,
   \ie $n_{i_1}= \ldots = n_{i_k} = n_j$:
   \begin{itemize}
   \item[]{\bf Case 2.2.1:}
   $2n_j \ne l$:
   \begin{equation}
    T(n_{i_1},\ldots,n_{i_k},n_j)
    \propto [T(n_{i_1},\ldots,n_{i_{k-1}}),T(n_{i_k},n_j)].
   \end{equation}
   \item[]{\bf Case 2.2.2:}
   $2n_j = l$; let $n_{i_k} = n'_{i_k} + n''_{i_k}$:
   \begin{equation}
    T(n_{i_1},\ldots,n_{i_k},n_j)
    \propto [T(n_{i_1},\ldots,n'_{i_k}),T(n''_{i_k},n_j)].
   \end{equation}
  \end{itemize}
 \end{itemize}
\end{itemize}

The cases include all possible variants and so the suggestion is
proved by recursion and all $l^n-1$ possible products of generators
except of $\Id$ can be represented using commutators.

\section{Universal set of quantum two-gates}

Elements $T_k$ have up to $n$ non-unit terms in tensor product \eq{genT2n}.
Here is described construction with no more than two terms. It is
used for description of a universal set of quantum two-gates and also
has direct analog with two-qubit gates.\cite{VlaUnCl}

Let us consider $B_0=T_0$, $B_j = T_j T_{j-1}^\hc$, $1 \le j < 2n$.
It is possible to generate full $\Tnl$ using the $2n$ elements:
$T_1 \propto [T_0,B_1]$, $T_i \propto [B_i,T_{i-1}]$, $\forall i>1$,
and so it is possible to generate recursively all $T_i$ and use
construction of $\Tnl$ described above.

Using \eq{genT2n} it is possible to write expressions for $B_j$:
\begin{mathletters}
\label{twoQG}
\begin{eqnarray}
 B_0 = T_0 & = & \Id^{\otimes (n-1)}\otimes T_x ,
 \label{twoQG:a}\\
 B_{2k+1}= T_{2k+1} T^\hc_{2k} & \propto &
 \Id^{\otimes (n-k-1)} \otimes T_z\otimes\Id^{\otimes k} ,
 \label{twoQG:b}\\
 B_{2k+2}= T_{2k+2} T^\hc_{2k+1} & \propto &
 \Id^{\otimes (n-k-2)}\otimes T_x\otimes T_x^\hc\otimes\Id^{\otimes k} ,
 \label{twoQG:c}
\end{eqnarray}
\end{mathletters}
with $k = 0,\ldots,n-1$ (or $n-2$).

To produce a universal set of quantum one- and two-gates it is enough
to use constructions of unitary matrices mentioned in the Introduction:
\begin{equation}
G^\tau_k = e^{i\tau(B_k + B_k^\hc)},\quad F^\tau_k = e^{\tau(B_k - B_k^\hc)}.
\label{TwoGates}
\end{equation}
It is possible to choose $\tau$ to express an arbitrary
matrix with given precision as product of matrices \eq{TwoGates}.

\renewcommand{\thefootnote}{\fnsymbol{footnote}}
\section*{Acknowledgment\protect\footnotemark[1]}
\footnotetext[1]{The acknowledgement and two new references are added only
in {\tt v2} (they are not reflected in JMP version),
further discussion about related topics can be found also in
E-prints: {\tt arXiv:quant-ph/0109010} and {\tt arXiv:math-ph/0112049}.}

Author is grateful to C. Zachos for comments to initial version of this paper.

\end{document}